\documentclass[prb,aps,amssymb,reprint,letterpaper,showpacs]{revtex4-1}

\usepackage{graphicx}
\usepackage{dcolumn}

\begin{document}

\title{Elastic properties and phase stability of AgBr under pressure.}

\author{P.T. Jochym}
\email{Pawel.Jochym@ifj.edu.pl}
\author{K. Parlinski}
\affiliation{Institute of Nuclear Physics,\\
  ul. Radzikowskiego 152, 31--342 Cracow, Poland}



\begin{abstract}
  {\it Ab initio} calculations have been used to derive the elastic
  constants, equation of state and free energy of NaCl, KOH and
  CsCl-type structures of AgBr.  The elastic constants have been
  derived by the stress--strain relation. The shear elastic constants
  C$_{44}$ has proved to be exceptionally soft. We have found that at
  temperature T=0 the phase transition from the low pressure NaCl-type
  to the high-pressure CsCl-type structure goes over a~monoclinic
  KOH-type structure.  The free energy comparison indicates that the
  monoclinic phase ranges from 8 to 35~GPa.
\end{abstract}


\maketitle


Silver halide crystals are of considerable interest, since they
exhibit deviation from the ideal ionic character due to the presence
of the d-electrons in Ag$^+$ ions. Numerous studies of these crystals
have been undertaken in the past years using various forms of
interaction potentials including three-body interactions and van der
Waals forces,\cite{singh,gupta} as well as the {\em ab initio\/} total
energy pseudopotential calculations.\cite{chelikowsky} There are also
many experimental works studying various aspects of these 
solids.\cite{loje,bridgmanA,bridgmanB,slykhouse,vaidya,fjeldly,mellander,ves}
The results of studies using phenomenological potentials generally
show good agreement with experiment.\cite{gupta} This studies use
sophisticated schemes involving three body interactions and van der
Waals potentials. The {\em ab initio\/} results for transition
pressure seems to be less accurate\cite{chelikowsky}.  There is one
{\em ab initio\/} study of phase stability of silver bromide
AgBr.\cite{nunes} Nevertheless, the transition mechanism is still
unclear for this compound. Several models have been proposed,
\cite{bridgmanA,bridgmanB,slykhouse,nunes,hull,kusaba,schock} but the
transition mechanism remains still an open problem. The {\em ab
  initio} study by Nunes, Allen and Martins,\cite{nunes} which
investigated the energy of several candidate structures, proposed an
intermediate trigonal phase with space group P$3_121$. In contrary
recent experimental work of Hull and Keen\cite{hull} suggested an
existence of an intermediate phase with a monoclinic, KOH-type
structure and shear instability as the transition mechanism.
The silver bromide AgBr is stable at ambient temperature and pressure
in the rocksalt structure Fm$\bar{3}$m.  In ambient temperature and
high pressures it has a~second stable phase with CsCl structure
Pm$\bar{3}$m.  The phase transition in between these phases has been
reported\cite{bridgmanB,slykhouse,bassett,riggleman} to occur at
P$_t$=8.3~GPa. In this paper we are studying elastic properties and
phase stability of the AgBr as a~function of pressure. For NaCl and
CsCl structures we have scanned pressures in the interval P=0--200~GPa,
and calculated the free energies and elastic constants. In the
range of P=0--40~GPa we have also studied the phase stability of the
free, non-constrained by any symmetry, AgBr crystal , searching for
any intermediate phases as well as general transition mechanism.  The
non-constrained crystal was treated as having P1 symmetry, however,
its unit cell content was not allowed to change. Hence, phase
transitions to phases with multiplication unit cells were not taken
into account.

\section{Methodology}

The calculations have been performed using Density Functional Theory
(DFT) approach with ultrasoft pseudopotentials and Generalised
Gradient Approximation (GGA).\cite{lin,goniakowski,payene} We have
used CASTEP\cite{castep} implementation of this method and
pseudopotentials provided with this package.  The elastic constants of
the NaCl phase have been also verified by the independent calculation
using VASP\cite{vasp}. The electronic minimisation used density mixing
scheme and pseudopotentials parametrised in the reciprocal space. The
summation over the Brillouin zone has been performed with weighted
summation over wave vectors generated by Monkhorst-Pack
scheme.\cite{pack} All computations were done in
$1$$\times$$1$$\times$$1$ supercell with 8 or 2 or 8 atoms for NaCl or
CsCl or KOH structures, respectively. Some tests for NaCl has been
carried out on the primitive, others on the face--centered cubic unit
cell.

Due to high demands of the accuracy of the free energy, we have
conducted extensive tests of convergence of the energy and pressure
with respect to energy cutoff and Brillouin zone grid spacing
parameters.  Thus, we have performed a~series of calculations with
cutoff energies in the range of $E_{cut}$=300~--~450eV and {\bf
  k}-space grid spacing in the range $\Delta${\bf
  k}=0.02~--~0.1~\AA$^{-1}$. In the light of obtained results we have
decided to use in our computations a~0.4~\AA$^{-1}$ grid spacing and
340eV energy cutoff. The chosen grid corresponds to 32 wave-vectors
for NaCl structure and 256 wave-vectors for CsCl structure. Such
a~large number of wave-vectors made calculations rather
time-consuming. But since we are comparing energies of {\em different}
structures with different unit cell volumes and Brillouin zone shapes,
sizes and {\bf k}-sampling, we need a very good convergence of energy.
At zero pressure the lattice constants of the optimised NaCl structure
is a= 5.7859~\AA, while the experimental lattice constants at P=0~GPa
and T=300~K is A=5.7745 \AA.

\section{Free Energy and Equation of State}

\begin{figure}[thbp]
  \includegraphics[angle=270,width=\columnwidth]{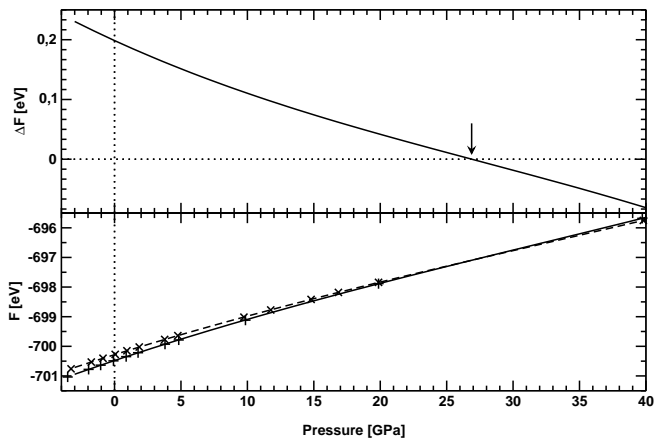}
  \caption{Lower part: Free energy of AgBr in NaCl 
    (+, solid line) and CsCl ($\times$, dashed line) structures. Upper
    part: Free energy difference F$_{\mathrm CsCl}-$G$_{\mathrm
      NaCl}$ between CsCl and NaCl structures as a~function of
    pressure. The vertical arrow denotes predicted phase transition
    pressure.}
  \label{fig:gibbs}
\end{figure}

\label{sec:gibbs}
To investigate the phase stability of the crystal and to estimate the
phase transition pressure, one must evaluate the free energy of all
involved phases and choose the one with the lowest energy as stable.
Since in {\em ab initio} calculations the temperature T=0, the free
energy is: $F=E+PV$. We have performed calculations of the crystal
free energy $F$ for several pressures from 0 to 50~GPa for both NaCl
and CsCl structures. Fig~\ref{fig:gibbs} shows the results of these
computations.  In the lower part the absolute values of free energies
per atom are plotted together with cubic polynomials fitted to these
data.  In the upper part the difference of $F_{CsCl}-F_{NaCl}$ is
plotted against pressure.  The difference changes sign at the pressure
P$_t$=27~GPa (arrow). We see that our calculations correctly predict
that the rocksalt structure is more stable at low pressure and that at
higher pressures the stability passes to CsCl structure .  The large
difference between experimental P$_t$=8.3~GPa and calculated
P$_t$=27~GPa transition pressure could be attributed to a few factors.
The main reason could be related to the appearance of the intermediate
KOH-type structure, discussed below.  Here, we mention other possible
factors.  First, the experimental data \cite{bridgmanA,bridgmanB} are
measured only at ambient temperature, while the calculations were
performed at T=0.  At finite temperature the entropy contribution to
the Gibbs free energy $G=E+PV-TS$ can also be different for the two
structures, since the closest coordination shells contain different
number of ions. Another reason could be related to the fact that the
van der Waals interaction is neglected in DFT calculations, and that
leads to under-binding of the crystal.  This effect has already
been noticed for silver iodide AgI crystal, where DFT predicts the
phase transition pressure to be 10~GPa,\cite{gupta,chelikowsky} while
experimentally it is 4~GPa.

\begin{figure}[thbp]
  \includegraphics[angle=270,width=\columnwidth]{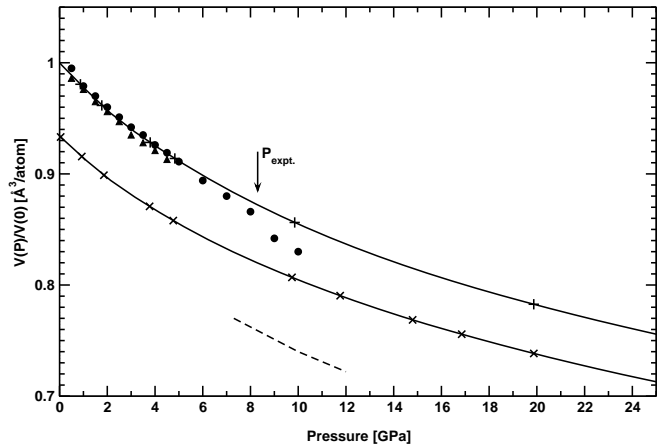}
  \caption{Variation of relative volumes with pressure 
    for NaCl(+) and CsCl ($\times$) structure. Experimental points
    ($\bullet,\blacktriangle$) are taken from Bridgman\cite{bridgmanA,bridgmanB} 
    and Vaidya\cite{vaidya}
    respectively.  The dashed line shows results of theoretical study
    of Gupta and Singh \cite{gupta}.
    The lines are present calculations.}
  \label{fig:volume}
\end{figure}

The data used to calculate free energy provide also the equation
of state of the crystals.  Fig.~\ref{fig:volume} shows the calculated
equation of state for NaCl and CsCl structures as well as the
experimental data measured by Bridgman and 
Vaidya.\cite{bridgmanA,bridgmanB,vaidya} We have also included in figure
\ref{fig:volume} the results of phenomenological study of Gupta and
Singh.\cite{gupta}  This work was able to predict the transition
pressure much better, but it fails to reproduce the volume compression
of the phase transition. Present calculations show excellent agreement
with experimental data for the low pressure range up to 6~GPa. In this
interval the relative error in V(P)/V(0) ratio is below 0.5\%.  It
should be noted that the Bridgman measurements, are limited to 10~GPa
and it would be interesting to increase the pressures to at least
15~GPa.  We have also fitted the Murnaghan equation of state to
rocksalt structure data below 15~GPa:
\begin{equation}
  \label{eq:murnaghan}
  V=V_0 \exp\left[-\frac{\ln\left(\frac{P A_2}{A_1}+1\right)}{A_2}\right]
\end{equation}
achieving good fit for $A_1$=41.01~GPa and $A_2$=4.35.

As seen from Fig.~\ref{fig:volume} the calculated relative volume change  
from NaCl to CsCl structure  at the
transition pressure P$_t$=8.3~GPa is a 6\% compression. We are not aware of
any experimental measurements of this value, however, we can estimate
it from the Bridgman data\cite{bridgmanB} to be at least 3\%.

\section{Elastic constants}

\begin{figure}[thbp]
  \includegraphics[angle=270,width=\columnwidth]{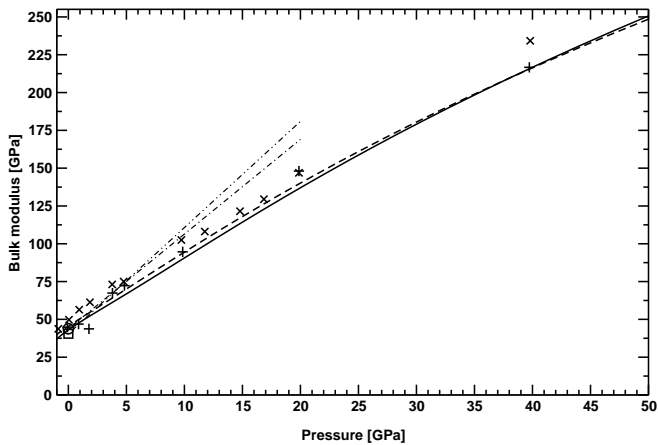}
  \caption{Bulk modulus of AgBr calculated for NaCl structure from 
    $P(V)$ function (solid line) and $c_{ij}$ data (+) from
    Fig.~\ref{fig:cij}. The bulk modulus for CsCl structure is
    calculated from $P(V)$ function (dashed line) and from $c_{ij}$
    data ($\times$) as well. Experimental data for T=195~K ($\circ$)
    and T=300~K ($\Box$) are taken from Loje\cite{loje}. The dotted lines
    denote experimental pressure derivatives of B at zero pressure.}
  \label{fig:b}
\end{figure}

The behaviour of elastic properties of the AgBr is specially interesting
 in the vicinity of the phase transition.  Fig.~\ref{fig:b}
shows pressure dependence of the bulk modulus as calculated from the
equation of state, as well as from the following relationship with elastic
constants:
\begin{equation}
  \label{eq:bdef}
  B=-V\frac{\partial P(V)}{\partial V}=\frac{1}{3}(c_{11}+2c_{12})
\end{equation}
The difference in the bulk modulus B between NaCl and CsCl structures
is quite small. On the same figure \ref{fig:b} we have included two
dotted lines showing experimental pressure derivatives of the bulk
modulus at P=0 and T=195~K and T=300~K.\cite{loje}  From these data we
see that our calculations reproduce the experimental bulk modulus
quite well and that the results from different methods are consistent
within 10\% .

We have used the standard stress--strain relationship to derive all
three elastic constants as a function of pressure. The method is based
on constructing a set of linear equations from stress--strain
relationships for several deformations of the unit cell. This set of
equations represents a general form of the Hook's law and can be
solved with respect to the elastic constants. Since in practice this
set of equations is over-determined, to solve it we have used a
singular value decomposition algorithm which automatically provides a
least squares solution of the set\cite{numrec}.  For each pressure
three negative and three positive deformations of each kind, namely,
bulk, shear and tetragonal elongation, have been set up.

\begin{figure}[thbp]
  \includegraphics[angle=270,width=\columnwidth]{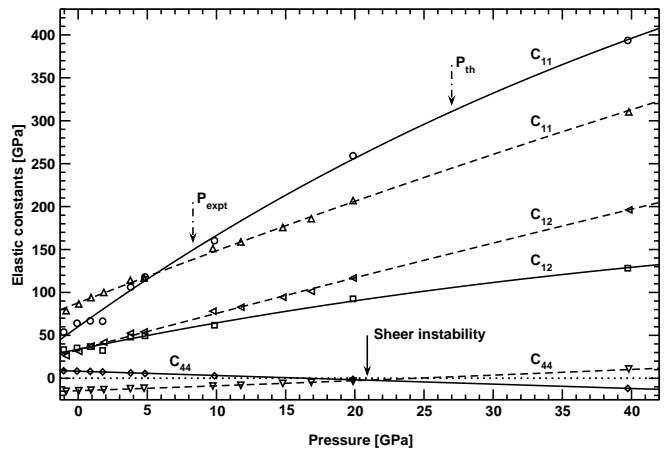}
  \caption{Elastic constants $c_{11}$, $c_{12}$, $c_{44}$ of 
    AgBr in NaCl (solid lines and $\circ$,~$\Box$,~$\diamond$) and
    CsCl (dashed lines and
    $\vartriangle,\vartriangleleft,\triangledown$) structures
    respectively.  The lines are cubic fits to the calculation data.
    The vertical solid arrow denotes predicted shear instability
    pressure.}
  \label{fig:cij}
\end{figure}

\begin{table}[b]
  \caption{Experimental and calculated bulk modulus $B$ and 
    elastic constants $c_{ij}$ of the AgBr crystal, in GPa.
    Values in parenthesis are calculated from 
    $B=\frac{1}{3}(c_{11}+2c_{12})$. Experimental data for T=195~K 
    and T=300~K are taken from Loje\cite{loje}.}
  \begin{ruledtabular}
  \begin{tabular}{ldddd}
    Result & 
    \multicolumn{1}{c}{$B$} & 
    \multicolumn{1}{c}{$c_{11}$} & 
    \multicolumn{1}{c}{$c_{12}$} & 
    \multicolumn{1}{c}{$c_{44}$} \\
    \hline
    Expt T=300~K
              & 40.5 & 56.1 & 32.7 & 7.2 \\
    Expt T=195~K 
              & 43.81 & 63.13 & 34.14 & 7.65 \\
    Present & (44.84) & 64.09 & 35.21 & 8.39 \\
    $B=-V\partial P(V)/\partial V$ &  
                 42.56  &       &       &       \\
  \end{tabular}
  \end{ruledtabular}
    \label{tab:elastic}
\end{table}

Table~\ref{tab:elastic} compares the calculated and experimental
elastic constants at ambient pressure and two temperatures: T=195~K
and T=300~K. Present results show quite good agreement with the
measurements, although the calculations are performed at T=0.  The
temperature data points allow to extrapolate the experimental values
to T=0. Thus, we have estimated that from T=195~K to T=0~K the elastic
constants might increase by about 5\%.  The mechanism of the phase
transition from NaCl to CsCl structure involves variation of the
$c_{44}$ elastic constant. We have calculated $c_{44}$ for both NaCl
and CsCl structures and for several pressures in the interval from 0
to 40~GPa . The results are shown at Fig.~\ref{fig:cij}.  We can
clearly see that the two lines representing the $c_{44}$ elastic
constants , intersect at pressure P=21~GPa, and moreover at value
close to zero.  The pressure P=21~GPa of the shear instability and
P=27~GPa of the free energy equality are not exactly the same.
However, we shall see in the next section that in this pressure
interval another phase could be more stable than the cubic NaCl or
CsCl structures.
\begin{figure}[b!]
  \includegraphics[width=\columnwidth]{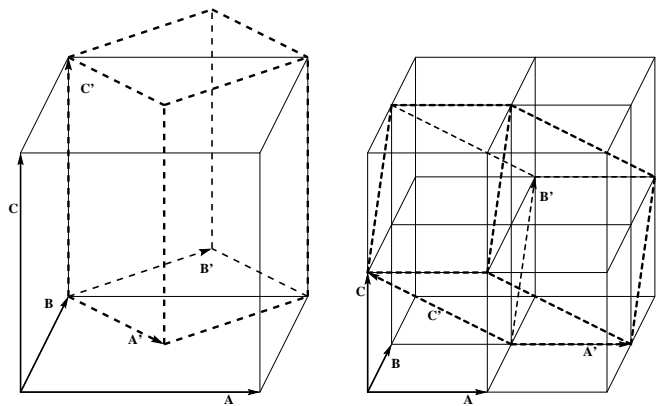}
  \caption{Relation between cubic NaCl and monoclinic KOH  (left) and 
    between cubic CsCl and monoclinic KOH (right) unit cells.
    The KOH unit cell is drawn with dashed lines.}
  \label{fig:struct}
\end{figure}
In previous sections the calculation were performed for NaCl and CsCl
structures under constrains that both phases remain always cubic.
Now, we release this restrictions and allow the unit cell to deform 
to a structure of P1 symmetry being intermediate configuration between
the NaCl and CsCl ones.

\section{Intermediate phase}

\begin{figure}[t]
  \includegraphics[angle=270,width=\columnwidth]{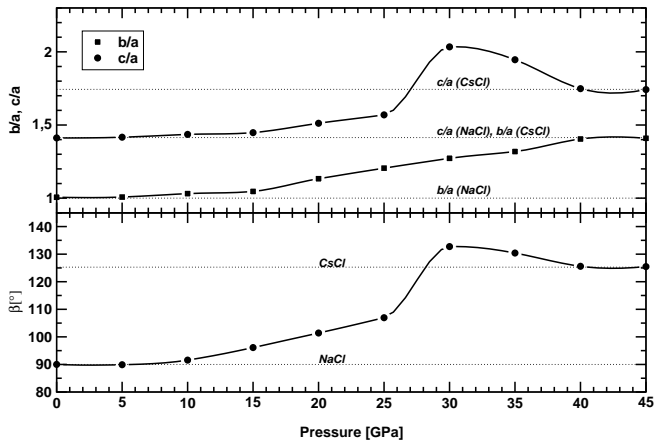}
  \caption{Transition from NaCl to CsCl structure. 
    Solid lines are guide to the eye.}
  \label{fig:jump}
\end{figure}
\begin{figure}[b!]
  \includegraphics[angle=270,width=\columnwidth]{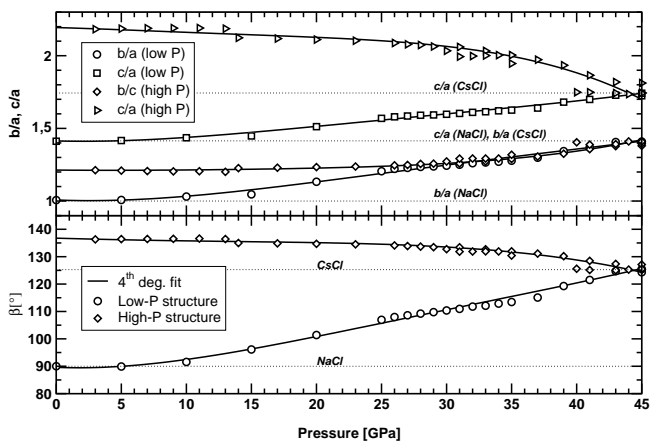}
  \caption{Cell parameters in NaCl structure to CsCl structure 
    transition of AgBr. 
    Solid lines are fourth degree fits to the data points.}
  \label{fig:cell}
\end{figure}
When the intermediate configuration becomes monoclinic it resembles
the KOH-type structure. The relationship between NaCl, KOH and CsCl
structures is shown on Fig.\ref{fig:struct}. Setting the monoclinic
angle $\beta$=90$^\circ$ or $\beta$=125,27$^\circ$ one reestablishes
the NaCl or CsCl structures, respectively. Hull and and
Keen\cite{hull} pointed out that this monoclinic phase, may provide a
proper transition path between NaCl and CsCl structures. A detailed
study of the above phase transition mechanism has been performed and
we refer the results below.  We have prepared a~model of the AgBr
crystal in the KOH-type structure with cell volume corresponding to
the NaCl structure at P=0. The $\beta$ angle of the monoclinic unit
cell was set to 110$^\circ$. Then the crystal geometry inducing atomic
relaxations was optimised with respect to free energy for various
increasing pressures from 0 to 45~GPa.  
\begin{figure}[t]
  \includegraphics[angle=270,width=\columnwidth]{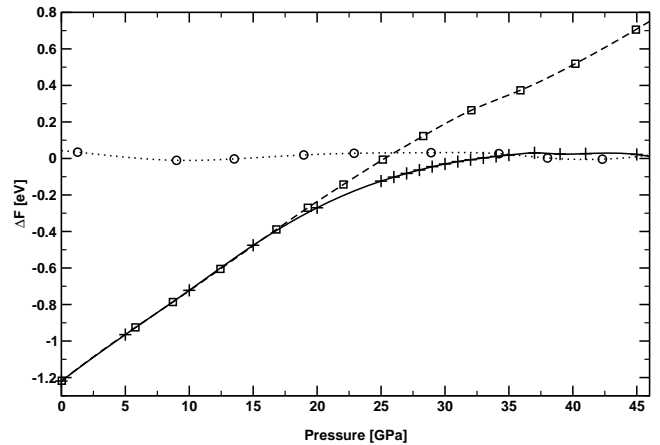}
  \caption{Free energy difference in the transition of AgBr from NaCl (squares)
    to CsCl structure (circles) through KOH structure (crosses). 
    The polynomial fit to the CsCl data points is taken 
    as the reference line (zero level). The lines are guide to the eye.}
  \label{fig:gibbs-koh}
\end{figure}
At each pressure the
minimisation procedure was reiterated several times since large
changes in the unit cell size involved changes of the Brillouin zone
grid.  The results are plotted in Fig \ref{fig:jump}.  One notices
that at about 8~GPa, there is a~clear tendency to depart from the pure
NaCl structure to an intermediate monoclinic structure.  There is also
a~clear jump of lattice parameters at the P=25--30~GPa range.
Simultaneously, the $\alpha$, and $\gamma$ angles do not change by
more then 0.05$^\circ$ for the scanned pressure range.  To verify the
last findings we have carried out two series of calculations each
starting from the same either low or high pressure configurations.
\begin{table}[b]
  \caption{Calculated elastic constants (in GPa) of the monoclinic, KOH-type 
    structure of the AgBr under pressure of P=20~GPa.}
  \begin{ruledtabular}
  \begin{tabular}{ddddddd}
    \multicolumn{1}{c}{Pressure [GPa]} &
    \multicolumn{1}{c}{$c_{11}$} & 
    \multicolumn{1}{c}{$c_{22}$} & 
    \multicolumn{1}{c}{$c_{33}$} & 
    \multicolumn{1}{c}{$c_{12}$} & 
    \multicolumn{1}{c}{$c_{13}$} & 
    \multicolumn{1}{c}{$c_{23}$} \\
    \hline
    20 & 156 & 175 & 229 & 152 & 84 & 98 \\
  \end{tabular}
  \end{ruledtabular}
  \label{tab:elmono}
\end{table}
The starting low and high pressure configurations were taken from
optimised monoclinic configurations at P$_{lo}$=25~GPa or
P$_{hi}$=30~GPa, respectively. The results are shown in
Fig.~\ref{fig:cell}. It is clear from this figure that both monoclinic
structures may show weak metastable behaviour. However, the energy
difference is small and it is possible that this effect will manifest
itself only in low temperatures. 
\begin{table*}[t!]
  \caption{Calculated structural parameters of the 
    intermediate KOH-type structure of the AgBr crystal.}
  \begin{ruledtabular}
  \begin{tabular}{ddddddd}
    \multicolumn{1}{c}{$P$ [GPa]} & 
    \multicolumn{1}{c}{$a$ [\AA]} & 
    \multicolumn{1}{c}{$b$ [\AA]} & 
    \multicolumn{1}{c}{$c$ [\AA]} &    
    \multicolumn{1}{c}{$\beta$ [$^\circ$]} &
    \multicolumn{1}{c}{$b/a$} &    
    \multicolumn{1}{c}{$c/a$} 
    \\
    \hline
    0 & 4.0803 & 4.1064 & 5.7601 &  90.0 & 1.006 & 1.412 \\
    5 & 3.9521 & 3.9819 & 5.5982 &  89.9 & 1.008 & 1.417 \\
    10 & 3.8226 & 3.9411 & 5.4886 &  91.6 & 1.031 & 1.436 \\
    15 & 3.7368 & 3.9091 & 5.4100 &  96.1 & 1.046 & 1.448 \\
    20 & 3.5445 & 4.0156 & 5.3589 & 101.4 & 1.133 & 1.512 \\
    25 & 3.4033 & 4.1011 & 5.3403 & 107.0 & 1.205 & 1.569 \\
    30 & 3.2993 & 4.1976 & 6.7106 & 132.7 & 1.272 & 2.034 \\
    35 & 3.2316 & 4.2612 & 6.2892 & 130.4 & 1.319 & 1.946 \\
    40 & 3.1725 & 4.4554 & 5.5465 & 125.6 & 1.404 & 1.748 \\
    45 & 3.1489 & 4.4375 & 5.4855 & 125.5 & 1.409 & 1.742 
  \end{tabular}
  \end{ruledtabular}
    \label{tab:struct}
\end{table*}
Furthermore, the energy of the
high-pressure configuration is always above the low-pressure one.
Consequently, the high-pressure structure is probably never present in
the real crystal, and we will omit this structure from now on.
Difference of the free energy of the three optimised candidate
structures (NaCl, KOH, CsCl) are shown in Fig.~\ref{fig:gibbs-koh}.
The polynomial fit to the CsCl free energy data is taken as the
reference line (``0'' level). At zero pressure the NaCl structure is the
most stable one. When we increase the pressure above about 8~GPa, the
cubic symmetry is broken, the monoclinic angle $\beta$ starts to
increase and the monoclinic phase is favoured. At the special point at
P=21~GPa, at which the shear elastic constants $c_{44}$ of NaCl and
CsCl structure vanish on the Fig.~\ref{fig:cij}, the monoclinic phase
is still the most stable one. Only above P=35~GPa the CsCl structure
becomes the most stable phase. However, for the finite temperature
case this might change, since the NaCl and CsCl structures have
different number of atoms in coordination shells and hence different
entropy terms in the Gibbs free energies.  Since there are no low
temperature measurements of pressure dependent behaviour of the AgBr we
cannot compare the present results with experimental findings.

We have also, for completeness, calculated some elastic constants of
the intermediate KOH phase of the crystal. The monoclinic phase has 13
independent elastic constants. However, due to the small energy
difference between phases connected by the shear deformation, only 6
of them can be calculated accurately. The remaining 7 constants are
too small and could be only estimated to be below 10~GPa. This is
quite expected result. The 6 large elastic constants, calculated under
pressure of P=20~GPa are presented in the Table~\ref{tab:elmono}. The
values are consistent with the values of elastic constants of NaCl and
CsCl structures depicted on the Fig.~\ref{fig:cij}.

Summarising, our results support the hypothesis formulated by Hull and
Keen,\cite{hull} that the phase transition from NaCl to CsCl phase,
goes through the intermediate KOH-type, monoclinic structure.

\section{Conclusions}

The present DFT calculations show that at temperature T=0~K the cubic
NaCl and CsCl-type structures of AgBr are stable at pressures 0--8~GPa
and above 35~GPa, respectively. Moreover, the calculation performed in
constrains of these two cubic structures permit to find the equation
of state and the elastic constants in the whole interesting range of
pressures from 0 to 45~GPa. It is interesting to see that the shear
elastic constants $c_{44}$ for both phases NaCl and CsCl vanish close
to 21~GPa. The derived equation of state and the elastic constants
characterise properly the stable cubic phase, and additionally can
provide an useful estimations of similar quantities within the
intermediate monoclinic phase. The calculation of the elastic
constants of the monoclinic KOH phase showed that, due to the small
energy difference between phases, the KOH phase is very soft to shear
deformations and its elastic constants are consistent with the elastic
constants of the other two phases.

The study of the intermediate phase has been done without point
symmetry elements constraints. This means that any deformation of the
simulated supercell could have taken place. The only imposed
limitation was the number of atoms in the supercell. Due to this
approach the supercell remained always monoclinic (or cubic for
special values of monoclinic angle $\beta$).  The values of the free
energy of the three candidate phases indicate that the monoclinic
phase becomes the most stable one in the pressure interval from 8 to
35~GPa.  This structure has previously been proposed to be the
transformation path between NaCl and CsCl-type phases. And indeed, the
free energy of the intermediate monoclinic phase seems to be lower
then these of the cubic ones. However the difference in energy is
quite small, which means that defects and impurities present in the
crystal could show a large impact on the monoclinic phase.

\begin{acknowledgments}
  The use of facilities of the ACC ``Cyfronet'', Cracow, where the
  calculations have been done, are kindly acknowledged.  This work was
  partially supported by the Polish State Committee of Scientific
  Research (KBN), grant No 2~PO3B~004~14, and computational grant No.
  KBN/SGI\_ORIGIN\_2000/IFJ/128/1998.
\end{acknowledgments}




\begin{thebibliography}{}
\bibitem{castep} CASTEP is a part of $Cerius^2$ system developed
  by Molecular Simulations.
\end{thebibliography}
\end{document}